\documentclass{emulateapj}

\shorttitle{{\sl{TRACE}} UV Wave Dependence on Magnetic Field}
\shortauthors{D. S. Bloomfield et~al.}

\begin{document}

\def \trace {{\sl{TRACE}}}
\def \soho {{\sl{SoHO}}}
\def \mdi {{\sc{mdi}}}
\def \ie {i.e.}
\def \eg {e.g.}

\title{The Influence of Magnetic Field on Oscillations in the Solar 
Chromosphere}

\author{D. Shaun Bloomfield\altaffilmark{1,2}, 
R. T. James McAteer\altaffilmark{3}, 
Mihalis Mathioudakis\altaffilmark{1}, 
and Francis P. Keenan\altaffilmark{1}}
\email{bloomfield@mps.mpg.de}
\altaffiltext{1}{Department of Physics and Astronomy, Queen's University 
Belfast, Belfast, BT7 1NN, Northern Ireland, U.K.}
\altaffiltext{2}{Max Planck Institut f\"{u}r Sonnensystemforschung, 37191 
Katlenburg--Lindau, Germany}
\altaffiltext{3}{NRC Research Associate, NASA / Goddard Space Flight Center, 
Solar Physics Branch, Code 682, Greenbelt, MD 20771, U.S.A.}

\begin{abstract}
Two sequences of solar images obtained by the {\emph{Transition Region and 
Coronal Explorer}} in three UV passbands are studied using wavelet and Fourier 
analysis and compared to the photospheric magnetic flux measured by the 
Michelson Doppler Interferometer on the {\emph{Solar Heliospheric 
Observatory}} to study wave behaviour in differing magnetic environments. 
Wavelet periods show deviations from the theoretical cutoff value and are 
interpreted in terms of inclined fields. The variation of wave speeds 
indicates that a transition from dominant fast--magnetoacoustic waves to slow 
modes is observed when moving from network into plage and umbrae. This implies 
preferential transmission of slow modes into the upper atmosphere, where they 
may lead to heating or be detected in coronal loops and plumes.
\end{abstract}


\section{INTRODUCTION}
\label{sec:intro}
It has been believed for over forty years that oscillatory motions play an 
important role in the outer solar atmosphere \citep{lei62}. This is 
particularly true of the wide variety of perturbations that are possible 
within the chromosphere, as this is the atmospheric layer which spans the 
transition from the domination of gas pressure to that of the magnetic field. 
These oscillations may be the signature of wave energy carried outward from 
the photosphere and deposited in the chromosphere, hence resulting in heating 
of the outer atmosphere. As such, spatial variations in oscillatory behaviour 
could yield information on the local topology of the magnetic environment 
\citep[\eg,][]{fin04}. \citet{fos05a} showed that hydrodynamic, 
high--frequency waves in the range $5-50$~mHz do not contribute sufficient 
energy to heat the atmosphere in a region of weak magnetic field. Also, 
\citet{soc05} found that electric currents in the chromosphere above a sunspot 
do not show a substantial spatial correlation with regions of either high 
temperature or large temperature gradient. Hence, it appears entirely 
plausible that differing forms of magnetoacoustic waves could supply some 
portion of the excess energy input through their existence in both extremes of 
the chromospheric environment.

Previous work on chromospheric oscillatory behaviour has reported on 
differences between spatial regions such as ``cell boundary'' and ``cell 
interior'' \citep[\eg,][]{dam84, deu90}, or the (equally as ambiguously 
defined) magnetic ``network'' and non--magnetic ``internetwork'' 
\citep[\eg,][]{lit93, mca04}. Although the findings of these authors are 
generally discussed in relation to the magnetic environment, spatial 
distinctions were often not based upon any magnetic field measurements, but 
instead on the magnitude of time--averaged emission within some spectral line 
or passband. This arbitrary form of classification may give the false 
impression that some sharp transition occurs when moving from the weak--field 
regime to that of higher field strengths. In reality, the observed 
distribution of solar magnetism shows a continuous variation of field 
strengths \citep{dom06}, suggesting that forms of oscillatory behaviour 
which are (in any manner) magnetically related should vary continuously as 
well \citep{law03}.

The comparison of chromospheric oscillations to surfaces of constant 
plasma--$\beta$, as determined by potential--field extrapolations from 
photospheric longitudinal magnetic fields \citep{mci03, mci04, fin04} has 
shown the spatial distribution of oscillatory power and propagation 
characteristics. We note that $\beta$, the ratio of gas to magnetic pressure, 
is given by $(2 / \gamma ) ( c_S / v_A )^2$, where $c_S$ is the sound speed 
(=$\gamma P / \rho$), $\gamma$ is the ratio of specific heats (=$5/3$), $P$ is 
the gas pressure, $\rho$ is the mass density, $v_A$ is the Alfv\'{e}n speed 
(=$B / \sqrt{4 \pi \rho}$), and $B$ is the magnetic field strength. However, 
the above studies mainly report upon the co--spatiality of oscillation 
behaviour with a chosen plasma--$\beta$ level (usually the unity surface; $c_S 
\approx v_A$) without yielding qualitative diagnostic feedback on the local 
atmospheric environment. In addition, the complicated two--dimensional 
characteristics of chromospheric wave propagation, both hydrodynamic 
\citep{ros02} and magnetoacoustic \citep{bog03} in nature, are only now 
beginning to be successfully explored through careful modeling. The results of 
such work are decidedly sobering, indicating a veritable plethora of 
interacting wave modes propagating at different speeds, sometimes in differing 
directions. Another consideration apparent from the simulations is the 
importance of knowing the plasma environment (\ie, gas or magnetic domination) 
in which the waves exist.

\begin{deluxetable*}{lccccccccc}
\tablecolumns{10}
\tablewidth{0pc}
\tablecaption{\trace\ UV and \mdi\ Data Set Summaries\label{tab:datasets}}
\tablehead{
\colhead{Data Set}  &  \colhead{Date}  &  \multicolumn{4}{c}{\trace}											      &  \multicolumn{4}{c}{\mdi}\\
		    &  		       &  \colhead{Start Time}	 &  \colhead{Frames}  &  \colhead{Cadence}  &  \colhead{Initial Pointing}		      &  \colhead{Start Time}	&  \colhead{Frames}  &  \colhead{Cadence}  &  \colhead{Initial Pointing}\\
		    &  		       &  \colhead{({\sc{UT}})}  &  		      &  \colhead{(s)}	    &  						      &  \colhead{({\sc{UT}})}  &		     &  \colhead{(s)}	   &  }
\startdata
1		    &  2004 Jul 03     &  15:50			 &  128		      &  16.0		    &  ($6^{\prime\prime}$, $-16^{\prime\prime}$)     &  15:37  		&  1	  	     &  \nodata		   &  ($-10^{\prime\prime}$, $-122^{\prime\prime}$)\\
2		    &  2000 Sep 22     &  08:00			 &  312		      &  22.7		    &  ($-174^{\prime\prime}$, $124^{\prime\prime}$)  &  08:01  		&  118  	     &  60.0		   &  ($-88^{\prime\prime}$, $44^{\prime\prime}$)
\enddata
\end{deluxetable*}

Our paper seeks to address the variable distribution of chromospheric UV 
intensity oscillations by studying their occurrence as a function of both 
detection height and the underlying photospheric magnetic flux. Observational 
data and their sources are discussed in \S~\ref{sec:data} with descriptions 
of the reduction and alignment processes, while \S~\ref{sec:anamet} outlines 
the wavelet and Fourier analysis. The time--localized nature of the wavelet 
transform is used to record the available distributions of wavepacket periods 
and durations. These parameters are of interest as oscillation periods can 
indicate which form of wave may exist (\ie, acoustic or magnetic) while the 
duration in terms of oscillatory cycles indicates the degree of energy carried 
if an oscillation is indeed the signature of a propagating wave. A Fourier 
phase difference technique is then used to directly study wave propagation 
characteristics. The results are presented and discussed in 
\S~\ref{sec:res_dis} in terms of wave behaviours and the plasma environment, 
while in \S~\ref{sec:conc} the implications of our work are summarized.

\section{DATA REDUCTION}
\label{sec:data}
\begin{figure}
\begin{center}
\plotone{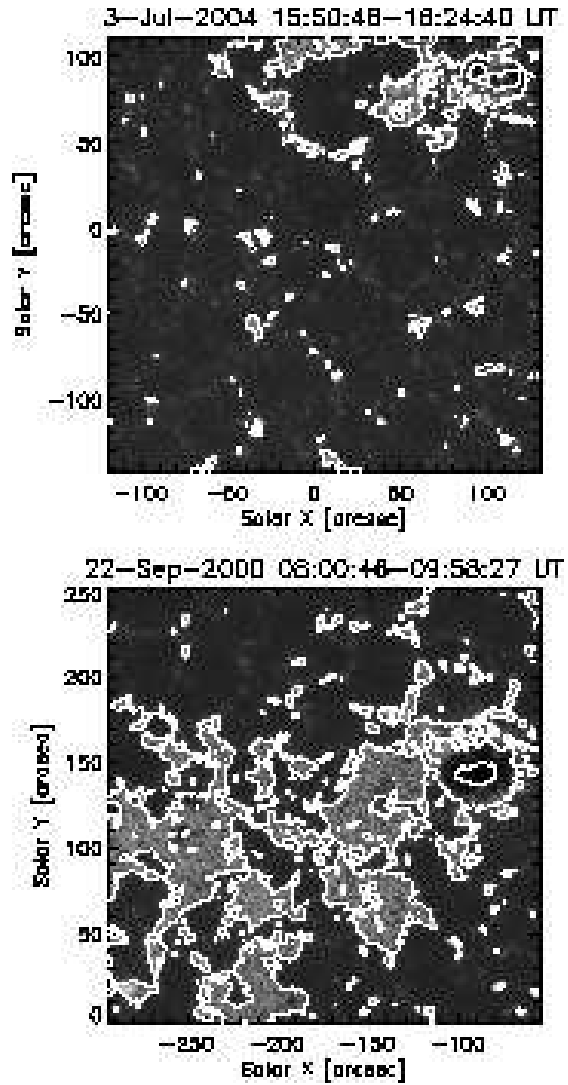}
\caption{\small Time--averaged \trace\ 1700~\AA\ images for set 1 
({\emph{upper}}) and set 2 ({\emph{lower}}). Overlaid thin (thick) contours 
mark the \mdi\ $\vert 50 \vert$~G ($\vert 500 \vert$~G) level.}
\label{fig:context}
\end{center}
\end{figure}

The chromospheric data studied were obtained by the {\emph{Transition 
Region and Coronal Explorer}} \citep[\trace;][]{han99}. Two $256 \times 
256$~arcsec$^2$ regions were recorded in the \trace\ UV passbands with a 
sampling of 0.5~arcsec~pixel$^{-1}$ on 2004 July 3 and 2000 September 22, with 
no compensation for solar rotation. Details of both data sets are given in 
Table~\ref{tab:datasets} while Figure~\ref{fig:context} depicts the 
time--averaged 1700~\AA\ emissions. In set 1, images were obtained in the 
order 1600~\AA, 1700~\AA, and 1550~\AA\ with a strict 4~s timing offset 
because of the \trace\ filterwheel set up, resulting in an overall stable 
cadence of 16~s for each individual passband, while set 2 comprised of 
successive 1550~\AA, 1600~\AA, 1700~\AA, and white--light images with an 
average cadence of 22.7~s.

Initially, data from each UV passband were dark subtracted and flat fielded. 
Portions of any images with corrupted data were replaced by the temporal mean 
of the preceding and following images, and JPEG compression effects were 
removed. Following \citet{def04}, data were derippled in both space and time 
by thresholding and removing spikes in the spatio--temporal Fourier domain. A 
final sweep of despiking was applied to remove cosmic ray hits and persistent 
``hot pixels'' were removed from the data. After this processing, each UV 
passband datacube was co--aligned by spatially cross correlating every image 
to the mid--image of the respective time sequence. Although difficult to 
attribute heights of formation (HOF) to the broadband UV emissions (due to the 
large range of temperature coverage involved in their production, and the 
large HOF overlap between the 1700~\AA\ and 1600~\AA\ passbands), we assumed 
that the HOF increased from the 1700~\AA\ passband to 1600~\AA\ and then to 
1550~\AA\ \citep{mca04, fos05b}. Mid--images of the 1600~\AA\ and 1550~\AA\ 
data were separately co--aligned to the mid--image of the 1700~\AA\ datacube 
for each data set, to achieve alignment between passbands. The 1700~\AA\ data 
were taken as the reference for the alignment process because this passband 
has the response function with the lowest formation height in the solar 
atmosphere out of the \trace\ passbands \citep[formed around the temperature 
minimum;][]{jud01}, thus allowing for a better reference comparison for the 
alignment to the underlying magnetic field data.

Photospheric line--of--sight (LOS) magnetic field data were obtained by the 
Michelson Doppler Interferometer \citep[\mdi;][]{sch95} on board the 
{\emph{Solar Heliospheric Observatory}} \citep[\soho;][]{fle95}. These data 
consist of longitudinal magnetic field measurements recorded with a spatial 
sampling of 0.6~arcsec~pixel$^{-1}$, as \mdi\ was operating in 
high--resolution mode. The \mdi\ magnetograms were processed to Level--1.8 by 
the standard data pipeline and compensated for the roll angle of the \soho\ 
spacecraft as well as the difference in image scales due to the differing 
observing positions of the satellites. In the case of set 2, magnetograms were 
co--aligned in the same manner as the UV datacubes before taking the temporal 
mean. Finally, \mdi\ pixels corresponding to the time--rotated \trace\ 
field--of--view (FOV) were re--sampled to the \trace\ pixel size. The 
correlation of the co--spatial portions of the two data sets is shown by the 
contours in Figure~\ref{fig:context}. This form of \mdi--to--\trace\ alignment 
was verified by spatially cross correlating the magnetic field magnitude to 
the intensity of the \trace\ 1700~\AA\ mid--image, achieving sub--pixel 
accuracy.

\begin{figure}
\begin{center}
\plotone{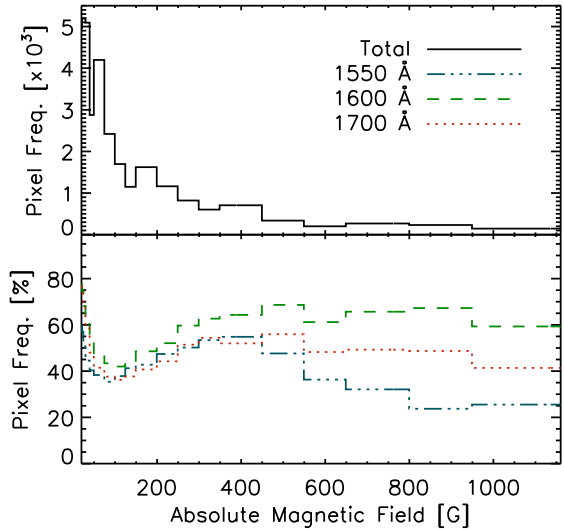}
\caption{\small {\emph{Upper}}: Frequency distribution of the total number of 
pixels ({\emph{ordinate}}) in the selected \mdi\ absolute field strength 
intervals ({\emph{abscissa}}) for set 1. {\emph{Lower}}: Percentages of the 
total number of pixels which show detections in each of the three \trace\ UV 
passbands. The distributions for detections in the 1700~\AA, 1600~\AA, and 
1550~\AA\ data are shown in dotted, dashed, and dotted--dashed lines, 
respectively.}
\label{fig:pixel_dists}
\end{center}
\end{figure}

It is important that the field strengths presented here are, as for any 
magnetogram instrument, interpreted as the integration of net flux over a 
pixel area. \citet{ber03} have shown, through direct comparison to the more 
reliable spectropolarimetry of the Advanced Stokes Polarimeter ({\sc{asp}}), 
that \mdi\ systematically underestimates the true longitudinal flux when 
operated in full--disk mode. However, we choose to report raw \mdi\ values 
because our work makes use of \mdi\ data obtained in high--resolution mode and 
the Berger \& Lites correction factors may not be accurate --- the difference 
in pixel sampling, and thus spatial flux averaging, should affect the slope of 
the \mdi--{\sc{asp}} flux--flux relation.

A magnetic field binning scheme was used in order to study oscillatory 
parameters in the \trace\ UV data as a function of the absolute \mdi\ magnetic 
field. Initially intervals of equal size were used, but the number of pixels 
decreases rapidly with increasing field strength. To counteract this, pixels 
with field strength values above $\vert 20 \vert$~G were instead separated 
into bins of systematically increasing size (listed in the first column of 
Table~\ref{tab:all_g_per}). The lower limit to the \mdi\ magnetic field 
strength intervals is directly based on the instrumental noise level being 
reported as $\sim$$\vert 20 \vert$~G \citep{sch95}. The probability 
distribution function (PDF) of pixels used in our work is illustrated for the 
case of set 1 in the upper panel of Figure~\ref{fig:pixel_dists}, while the 
lower panel depicts the percentage of pixels that contain wavelet oscillation 
detections.

\section{ANALYSIS METHODS}
\label{sec:anamet}

\subsection{Wavelet Power Transform}
\label{subsec:wavpowtran}
One of the most prevalent tools that has been recently used in observational 
studies of oscillatory behaviour is the technique of wavelet analysis (see the 
excellent introduction provided by Torrence \& Compo 1998 and its application 
to solar data in, \eg, De Moortel et al. 2000, McAteer et al. 2004, and 
McIntosh \& Smilie 2004). This multiscale analysis is better suited than 
Fourier analysis to both the spatial and temporal non--stationarity exhibited 
by oscillations in the solar atmosphere. For our study of oscillatory signals 
we chose the standard Morlet wavelet, a Gaussian--modulated sine wave which 
has an associated Fourier period, $P$, corresponding to $1.03s$, where $s$ is 
the wavelet scale. The wavelet transform is a convolution of the time series 
with the analyzing wavelet function whereby the complete wavelet transform is 
achieved by varying the wavelet scale, which controls both the period and 
temporal extent of the function, and scanning this through the time series. 
Regions exist at the beginning and end of the wavelet transform where spurious 
power may arise as a result of the finite extent of the time series. These 
regions are referred to as the cone of influence (COI), having a temporal 
extent equal to the $e$--folding time of the wavelet function. For the case of 
the Morlet wavelet it has the form, $t_d = \sqrt{2} s = \sqrt{2} P / 1.03$. 
This time scale is effectively the response of the wavelet function to noise 
spikes and is used in our detection criteria by requiring that oscillations 
have a duration greater than $t_d$ outside the COI. This imposes a maximum 
period above which we can not accept detections, since we are dealing with 
finite time series of duration $D_{tot}$. At this maximum period it follows 
that $P_{max} = 1.03 D_{tot} / 3 \sqrt{2}$ --- for set 1 $D_{tot}$ is 2032~s 
and $P_{max}$ is 493~s, while $D_{tot}$ is 7060~s and $P_{max}$ is 1714~s for 
set 2.

The form of automated wavelet analysis carried out on these data has 
previously been presented in \citet{mca04}, although some revisions have been 
carried out for our work. Firstly, no filtering was applied to data set 1 
because the duration of observation presented here (34~min) is small compared 
to that of the \trace\ spacecraft orbital period ($\sim$$90$~min). Therefore, 
long--period variations due to changes in Earth upper--atmospheric 
transmission are of small amplitude when compared to shorter periods of solar 
origin. Secondly, the more stringent statistical criterion of oscillations 
being at least 99\% confident is employed rather than previous values of 95\%. 
Finally, all power in the transform is retained, even that in the COI. Any 
detection having power above the 99\% level for at least $t_d$ outside the COI 
has the entire duration recorded, including any portion which falls in the 
COI. This modification helps somewhat to circumvent previous limits on 
durations of detections. Not only does the upper limit to the possible number 
of cycles decrease with increasing oscillatory period (since analyzing finite 
duration time series), but removing the COI decreases this upper limit more 
drastically as the COI both increases with period and occurs at the start and 
end of the wavelet transform. After a complete analysis, carried out on the 
time series of each spatial pixel in all three \trace\ passbands, the output 
comprises the duration in oscillatory cycles and the period of every detection.

\subsection{Fourier Phase Differences}
\label{subsec:fft_phdiff}
A Fourier phase difference analysis was used to further study possible 
propagation characteristics of waves between the co--spatial 1700~\AA\ and 
1600~\AA\ time series. These passbands were chosen based on the proximity of 
their peak formation heights \citep{fos05b}. This form of Fourier analysis is 
a technique which allows investigation of the difference in waveforms between 
two time series. Using the phase information contained within the complex 
Fourier transform, via the equations discussed in detail by \citet{kri01}, the 
difference in cyclic phase, $\Delta \phi$, can be determined for each Fourier 
frequency component, $\nu$, with the quality of the values represented by the 
phase coherence. The pairs of 1700~\AA\ and 1600~\AA\ time series in this 
study are extracted from the same ($x$, $y$) pixel location, but the signals 
remain separated in the direction normal to the solar surface. In this 
scenario, phase differences can be interpreted as delays caused by the finite 
propagation speed of waves traveling between the formation heights. The 
1700~\AA\ data were linearly interpolated to the 1600~\AA\ 
(equally--separated) times to compensate for timing offsets that result from 
the \trace\ UV data acquisition (\S~\ref{sec:data}). These offsets require 
removal as they produce drifts in phase difference spectra with frequency 
\citep{kri01}.

\begin{figure}
\begin{center}
\plotone{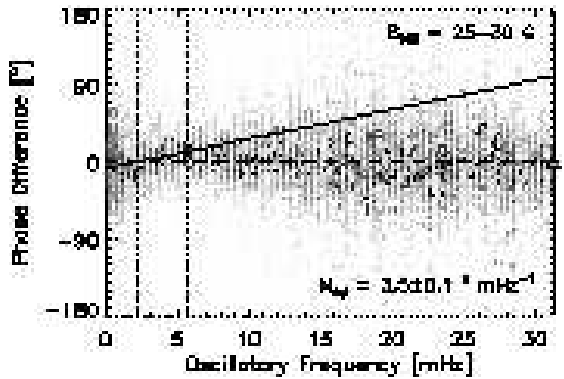}
\caption{\small Fourier phase differences ({\emph{ordinate}}) between 2420 
pairs of \trace\ 1700~\AA\ and 1600~\AA\ time series as a function of 
oscillatory frequency ({\emph{abscissa}}). Larger symbols and darker shading 
depicts increased weighting significance (\ie, increased Fourier cross--power 
and coherence). The phase difference gradient ($M_{\Delta \phi}$; solid line) 
was calculated by weighted least--squares fitting of a first--order polynomial 
to all points in the range $2.0-5.4$~mHz, indicated by vertical dotted lines.}
\label{fig:fft_exam}
\end{center}
\end{figure}

At those frequencies over which a wave has a real solution to its dispersion 
relation, and may thus propagate, phase difference spectra between emission 
signals formed at two separate heights are expected to have the form,
\begin{equation}
\label{eqn:ph_grad}
\Delta \phi \left( \nu \right) \approx \frac{\Delta h}{v_{ph}} \nu \ ,
\end{equation}
where $\Delta h$ is the separation distance of the emitting regions, $v_{ph}$ 
is the phase speed of the wave, and $\nu$ is the oscillation frequency. In a 
method similar to \citet{mci03}, phase difference gradients ($M_{\Delta \phi} 
\approx \Delta h / v_{ph}$) were extracted over the frequency range 
$2.0-5.4$~mHz (periods $185-500$~s). This range was chosen because it 
encompasses the majority of our oscillation detections 
(\S~\ref{subsec:per_bmdi}). As shown in Figure~\ref{fig:fft_exam}, phase 
difference gradients were achieved by applying a weighted least--squares fit 
of a first--order polynomial to the phase difference spectra obtained from all 
time series originating from pixels in a particular range of field strengths. 
The contribution of each phase difference point to the fit was weighted by 
both the Fourier cross--power and coherence values.

\section{RESULTS \& DISCUSSION}
\label{sec:res_dis}
Wavelet results are presented only for set 1 to avoid repetition, since set 2 
displays similar behaviour with longer--period detections due to the increased 
duration of observation and more short--period detections from flare surges in 
the time series. However, Fourier results from both data sets are provided and 
discussed.

\subsection{Wavelet Period Distributions}
\label{subsec:per_bmdi}
\begin{figure}
\begin{center}
\plotone{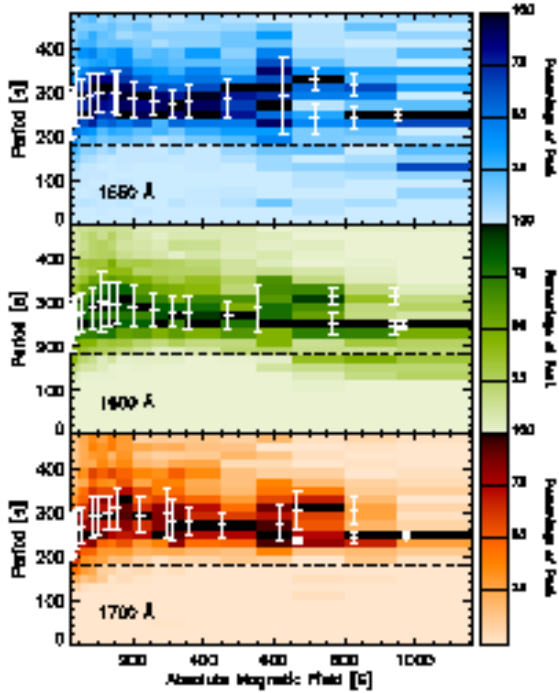}
\caption{\small Frequency distributions of \trace\ 1700~\AA\ ({\emph{lower}}), 
1600~\AA\ ({\emph{middle}}), and 1550~\AA\ ({\emph{upper}}) wavelet periods 
({\emph{ordinate}}) as a function of the \mdi\ photospheric magnetic flux 
magnitude ({\emph{abscissa}}). Distributions are normalized to their peak 
number of detections. Symbols mark centroid positions of Gaussian fits to each 
period distribution, with bars extending over the $\pm$1$\sigma$ width 
(detailed in Table~\ref{tab:all_g_per}). Points are overlaid at the mean \mdi\ 
value of pixels showing detections in that field strength range. The dashed 
line marks the non--magnetic value of $P_{\mathrm{ac}}$ for $\theta = 0$ at 
the temperature minimum.}
\label{fig:perbmagall}
\end{center}
\end{figure}

Detections in each \mdi\ field strength interval were sampled 
in terms of their period with uniform $20$~s binning. The resulting 1700~\AA, 
1600~\AA, and 1550~\AA\ PDFs are shown in Figure~\ref{fig:perbmagall}, where 
the individual distributions are self--normalized to their peak number of 
detections to highlight changes between PDFs by compensating for the decreased 
number statistics at increased field strengths (Figure~\ref{fig:pixel_dists}). 
Fitting profiles which consisted of a single Gaussian and a linearly 
increasing first--order polynomial (as most distributions showed a tail of 
detections at high periods) were applied to each PDF. Although the 
distributions are not ideally Gaussian, the variation of fitted centroid 
period ($P_c$; displayed as central bar marks in Figure~\ref{fig:perbmagall} 
and detailed in Table~\ref{tab:all_g_per}) with \mdi\ field strength and UV 
passband formation height should at least qualitatively describe these data. 
Bars extend over the $\pm$1$\sigma$ width of the Gaussian fit in 
Figure~\ref{fig:perbmagall}, representing PDF spread and not error in the fit. 
For each UV passband the symbols are plotted at the mean absolute field 
strength of the pixels which show oscillations in that passband. For a 
particular range of \mdi\ field strengths the mean value will vary between 
each of the three UV passbands, a direct result of the differing pixel numbers 
which contribute to the different UV passband oscillation distributions (lower 
panel of Figure~\ref{fig:pixel_dists}).

Period PDFs in the range $\vert 25-100 \vert$~G shift to higher values of 
$P_c$ when moving to larger \mdi\ field strengths. This progression of periods 
in the low chromosphere from 4-- to 5-- min agrees with the findings of 
\citet{law03} who analyzed 3~\AA--wide passband observations of \ion{Ca}{2}~K 
line intensity in the quiet--Sun (inter)network. Although the K$_2$ portion of 
the \ion{Ca}{2}~K line profile is formed at least somewhat above the 
temperature minimum (\ie, above these \trace\ data), close 
agreement may still be expected with this work because the K$_1$ wing emission 
also contained in their broad passband is formed around the 
temperature minimum \citep{ver81}. Lawrence et al. attribute the period change 
to their passband being formed below the $\beta = 1$ level at low field 
strengths and above it for higher fields, without offering a physical reason 
for the change. We draw attention to the fact that the periods detected are 
incompatible with standard MHD theory, 
where waves can only propagate at periods below the acoustic cutoff 
($P_{\mathrm{ac}} \propto \sqrt{T}$; $T$ is temperature). As such, waves up to 
$\sim$5~min period can propagate into the photosphere (\eg, $p$ modes) but the 
temperature decrease with height in the low chromosphere acts to decrease the 
cutoff period and previously propagating long--period waves become evanescent.

We extend the hypothesis of \citet{law03} by proposing that wave propagation 
may be permitted and the period behaviour explained using both the direct 
dependence that $P_{\mathrm{ac}}$ has on the magnetic field strength 
\citep{bel71} and the inclination, $\theta$, which arises through the 
effective gravity, $g$ 
\citep[$P_{\mathrm{ac}} \propto 1 / g = 1 / g_0 \cos \theta$, where $g_0 = 274$~m~s$^{-1}$;][]{depon04}. 
As shown by \citet{law03}, regions of increased magnetic field experience the 
$\beta = 1$ level at lower altitudes. This intuitively suggests that the 
``magnetic canopy'', where the field spreads nearly horizontally, will be 
formed lower. If for low field strengths a HOF lies significantly below the 
``canopy'' level it will experience near--vertical field lines. However, as 
the photospheric flux increases, the $\beta = 1$ and ``canopy'' levels will be 
pulled down toward the HOF such that it samples increasingly inclined fields. 
The larger inclination acts to reduce the effective gravity, which in turn 
increases the cutoff period allowing for ``tunneling'' of evanescent wave 
modes \citep{depon04}. Over the range $\vert 22-75 \vert$~G, increasing values 
of $P_c$ are recorded when moving upward through the UV passband HOFs, 
agreeing with this picture of sampling field lines of increasing inclination 
with height.

\begin{deluxetable}{ccccccc}
\tablecolumns{7}
\tablewidth{0pc}
\tablecaption{Magnetic Field Strength Intervals and Fit Parameters to Wavelet 
Period Distributions of 2004 July 03 (Set 1)\label{tab:all_g_per}}
\tablehead{
\colhead{\mdi\ Field}	&  \multicolumn{2}{c}{1700~\AA\ Period}		&  \multicolumn{2}{c}{1600~\AA\ Period}		&  \multicolumn{2}{c}{1550~\AA\ Period}\\
\colhead{(G)}		&  \multicolumn{2}{c}{(s)}			&  \multicolumn{2}{c}{(s)}			&  \multicolumn{2}{c}{(s)}\\
			&  \colhead{$P_c$}  &  \colhead{1$\sigma$}	&  \colhead{$P_c$}  &  \colhead{1$\sigma$}	&  \colhead{$P_c$}  &  \colhead{1$\sigma$}}
\startdata
$\vert 20-22 \vert$     &  238.3  &  $\pm 47.1$  &  236.6  &  $\pm 49.6$  &  242.7  &  $\pm 52.6$\\
$\vert 22-25 \vert$     &  238.6  &  $\pm 46.4$  &  239.0  &  $\pm 48.8$  &  244.0  &  $\pm 49.1$\\
$\vert 25-30 \vert$     &  243.9  &  $\pm 48.8$  &  245.2  &  $\pm 53.3$  &  250.2  &  $\pm 50.0$\\
$\vert 30-40 \vert$     &  253.8  &  $\pm 47.8$  &  257.6  &  $\pm 48.4$  &  266.7  &  $\pm 45.9$\\
$\vert 40-50 \vert$     &  261.2  &  $\pm 46.0$  &  268.3  &  $\pm 44.7$  &  288.6  &  $\pm 66.6$\\
$\vert 50-75 \vert$     &  270.0  &  $\pm 41.2$  &  276.6  &  $\pm 45.8$  &  284.8  &  $\pm 43.6$\\
$\vert 75-100 \vert$    &  289.9  &  $\pm 48.8$  &  288.5  &  $\pm 47.8$  &  293.6  &  $\pm 49.6$\\
$\vert 100-125 \vert$   &  293.6  &  $\pm 42.5$  &  302.7  &  $\pm 66.8$  &  300.4  &  $\pm 44.8$\\
$\vert 125-150 \vert$   &  298.3  &  $\pm 39.6$  &  296.7  &  $\pm 49.8$  &  300.9  &  $\pm 48.5$\\
$\vert 150-200 \vert$   &  310.0  &  $\pm 47.1$  &  298.4  &  $\pm 48.8$  &  297.6  &  $\pm 50.3$\\
$\vert 200-250 \vert$   &  294.0  &  $\pm 40.3$  &  290.1  &  $\pm 47.2$  &  283.5  &  $\pm 38.7$\\
$\vert 250-300 \vert$   &  298.5  &  $\pm 58.7$  &  282.1  &  $\pm 38.3$  &  279.3  &  $\pm 31.6$\\
$\vert 300-350 \vert$   &  282.9  &  $\pm 44.5$  &  278.7  &  $\pm 36.4$  &  275.3  &  $\pm 32.8$\\
$\vert 350-450 \vert$   &  279.9  &  $\pm 31.2$  &  278.1  &  $\pm 36.0$  &  279.8  &  $\pm 35.5$\\
$\vert 450-550 \vert$   &  270.7  &  $\pm 25.8$  &  270.3  &  $\pm 32.0$  &  286.2  &  $\pm 41.4$\\
$\vert 550-650 \vert$   &  273.9  &  $\pm 40.9$  &  287.1  &  $\pm 54.4$  &  293.0  &  $\pm 87.1$\\
$\vert 650-800 \vert$   &  236.4  &  $\pm  8.2$  &  252.0  &  $\pm 23.7$  &  239.6  &  $\pm 33.9$\\
\nodata			&  307.0  &  $\pm 43.9$  &  314.0  &  $\pm 18.3$  &  329.5  &  $\pm 25.0$\\
$\vert 800-950 \vert$   &  243.3  &  $\pm 12.8$  &  243.7  &  $\pm 16.0$  &  242.4  &  $\pm 25.9$\\ 
\nodata			&  305.4  &  $\pm 33.5$  &  311.9  &  $\pm 18.7$  &  316.1  &  $\pm 25.6$\\
$\vert 950-1200 \vert$  &  245.7  &  $\pm  6.1$  &  247.9  &  $\pm  9.9$  &  248.3  &  $\pm 13.2$\\
$> \vert 1200 \vert$    & \nodata &  \nodata     & \nodata &  \nodata     & \nodata &  \nodata
\enddata
\end{deluxetable}

All three UV passbands subsequently show decreasing centroid periods when 
moving into regions of increased magnetic field ($\vert 125-550 \vert$~G). 
This differs from the results of \citet{law03} where the peak wavelet power 
was observed at a fairly constant 5--min period for field strengths in the 
range $\vert 200-500 \vert$~G, although a marginal downturn in period may 
exist at their highest fields. However, it should be noted that the solar 
environment studied by \citet{law03} contained neither plage nor sunspot 
regions which contribute to those field strengths in this work. A more 
appropriate comparison can be made with data set 2 previously presented in 
\citet{mci04} which is our set 2. Although not discussed, comparing the 
detected wavelet frequency in their Fig.~5 with the magnetic field shows the 
same variation: short period oscillations in the internetwork, increasing 
periods in areas of network and outer plage, and a decrease in period in 
those plage regions corresponding to the strongest plage magnetic fields.

The \mdi\ intervals containing all but the largest field strengths ($\vert 
650-950 \vert$~G) exhibit double--peaked PDFs. This makes qualitative sense 
as a sunspot lies in the FOV. As such, the lower--period components 
contain the $\sim$3--min umbral oscillations \citep[][more distinct in data 
set 2 for fields $> \vert 1200 \vert$~G where periods peak at 
$160-180$~s]{gio72}, while components at higher periods contain the 
$\sim$5--min penumbral oscillations \citep{bec72, moo75}. It should be noted 
that period PDFs for all passbands do not appear double peaked at the highest 
fields ($> \vert 950 \vert$~G), despite showing decreased number statistics. 
The disappearance of a higher--period component at these largest field 
strengths supports their interpretation as penumbral oscillations since these 
fields should sample only umbral regions.

\subsection{Wavelet Duration Distributions}
\label{subsec:cyc_bmdi}
\begin{figure}
\begin{center}
\plotone{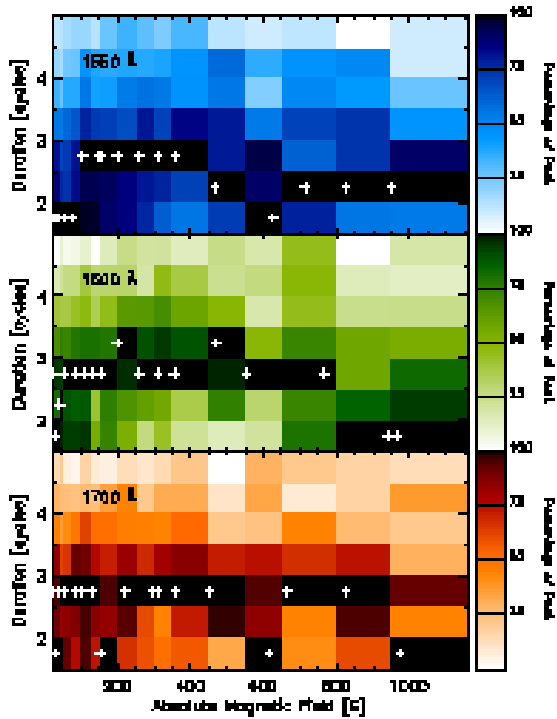}
\caption{\small As Fig.~\ref{fig:perbmagall}, but for \trace\ 1700~\AA\ 
({\emph{lower}}), 1600~\AA\ ({\emph{middle}}), and 1550~\AA\ ({\emph{upper}}) 
durations of wavelet detection ({\emph{ordinate}}). Symbols mark the most 
frequently detected duration, plotted at the mean field strength of pixels 
showing oscillations in that field strength range.}
\label{fig:cycbmagall}
\end{center}
\end{figure}

An alternate way by which to characterize oscillations is their duration of 
observation, since this quantity can be interpreted as a measure of the energy 
associated with a wave. As such, detections were sampled in terms of their 
duration with uniform half--cycle binning. The resulting PDFs are shown 
separately in Figure~\ref{fig:cycbmagall} for the 1700~\AA, 1600~\AA, and 
1550~\AA\ data, where every PDF is again normalized to its peak number of 
detections. In these plots the most frequently detected duration is marked by 
a `$+$' symbol, and differing dependencies on the magnetic field are clearly 
seen. The 1700~\AA\ data have oscillations of duration $2.5-3$~cycles most 
frequently observed across the majority of magnetic fields, while the 
1600~\AA\ data also peak at $2.5-3$~cycles over most field strengths. The 
large degree of similarity between the behaviours of these two sets of 
distributions strongly implies that both UV passbands are formed in 
essentially the same physical environment. However, the 1550~\AA\ durations 
peak at $1.5-2$~cycles over low field strengths ($\vert 25-100 \vert$~G) 
before changing to $2.5-3$~cycles for moderate magnetic fields ($\vert 100-450 
\vert$~G). The oscillation durations then shorten to $2-2.5$~cycles for nearly 
all higher field strengths.

It appears that, if these oscillations are indeed waves, below $\vert 100 
\vert$~G some portion of the wave energy budget has been lost between the 
formation heights of 1600~\AA\ and 1550~\AA, since the 1550~\AA\ data exhibit 
durations shorter than both the 1700~\AA\ and 1600~\AA\ data. By contrast, the 
wavelet detection durations are conserved in the stronger--field range $\vert 
100-450 \vert$~G which, in turn, indicates that the effective ``boundary'' 
previously lying between the 1600~\AA\ and 1550~\AA\ HOFs has been lowered in 
altitude to below the 1700~\AA\ HOF. Incorporating a potential--field 
extrapolation, \citet{law03} interpret their change from 4-- to 5-- min 
periods at field strengths of $\sim$$\vert 150 \vert$~G as resulting from the 
lowering of the $\beta$--unity surface below the formation height of their 
\ion{Ca}{2}~K line emission. This is very close to our wavelet duration 
transition field strength of $\vert 100 \vert$~G, while the disparity may be 
explained by the difference in passband formation heights and/or the 
assumption of the static, spatially--averaged VAL C model atmosphere 
\citep{ver81} in the extrapolation.

\subsection{Fourier Phase Speeds}
\label{subsec:fft_phvel}
\begin{figure}
\begin{center}
\plotone{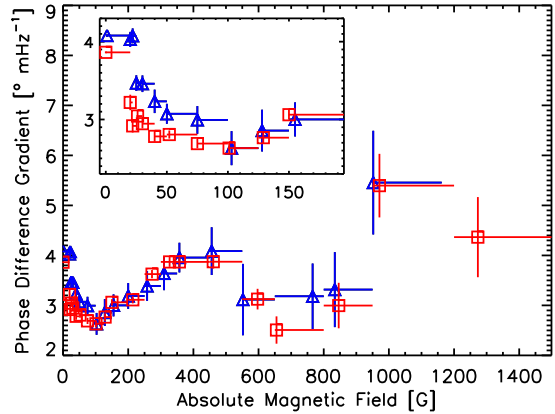}
\caption{\small Fourier phase difference gradients ($M_{\Delta \phi}$; 
{\emph{ordinate}}) between the 1700~\AA\ and 1600~\AA\ intensity time series 
as a function of \mdi\ magnetic flux magnitude ({\emph{abscissa}}) --- 
triangles correspond to set 1 and squares to set 2. Symbols are plotted at the 
mean absolute field strength value, where horizontal bars denote the extent of 
the field strength range and vertical bars mark the $\pm$1$\sigma$ uncertainty 
to the weighted least--squares linear fits. The values obtained over 
low--field values are displayed within the inset to highlight fine detail.}
\label{fig:fft_grad}
\end{center}
\end{figure}

Although highly versatile and a powerful tool for studying oscillatory 
behaviour, the wavelet analysis performed up to now has not been able to 
distinguish whether the detected oscillations are the signatures of 
propagating waves in the solar atmosphere. It is for this reason that Fourier 
phase difference analysis was applied to the same data. The results of the 
analysis, outlined in \S~\ref{subsec:fft_phdiff}, are presented in 
Figure~\ref{fig:fft_grad} for both data sets as the weighted--fit values of 
phase difference gradient between the 1700~\AA\ and 1600~\AA\ time series 
pairings.

If a spatially--invariant atmosphere is assumed, the vertical separation of 
UV HOFs will not vary when moving into regions of increased magnetic field. As 
a result, $\Delta h$ is constant and Equation~\ref{eqn:ph_grad} informs us 
that a change in Fourier phase difference gradient denotes the inverse of the 
change that has occurred in the wave phase speed. In the past, such forms of 
atmosphere have been used to implement potential--field extrapolations 
\citep[\eg,][]{mci03, law03}. Unfortunately, these act to oversimplify the 
complexity of the magnetic atmosphere and the application of 1--D models that 
take account of the differences between regions of differing magnetic field 
will yield more appropriate results.

From consideration of Equation~\ref{eqn:ph_grad}, it is possible that the 
initial decrease of gradient values in Figure~\ref{fig:fft_grad} could result 
from waves having the same phase speed but the HOF separations are reduced 
with increasing magnetic field. This suggests that $M_{\Delta \phi}$ should 
continue decreasing for even stronger fields, but we detect a switch in 
behaviour to increasing $M_{\Delta \phi}$ in the range $\vert 125-550 \vert$~G 
which requires increasing separations between the HOFs. This change in 
separation behaviour does not make qualitative sense, so we proceed by 
assuming a spatially--invariant atmosphere that yields increasing wave phase 
speeds up to $\vert 100-125 \vert$~G and a subsequent decrease in speeds up to 
$\vert 550 \vert$~G. A decrease in \trace\ UV phase difference gradient with 
increasing magnetic field has been previously observed by \citet{mci03} with 
no turnaround reported. The deviation of our results can be explained by their 
quiet--Sun (inter)network data having fields $\le$$\vert 50 \vert$~G so the 
$\beta = 1$ level was not lowered below the UV HOFs, while wavelet durations 
imply that this occurs for fields $\ge$$\vert 100 \vert$~G in this study.

The observed wave speed behaviour over low-- to moderate-- magnetic fields may 
be interpreted as resulting from a change in physical environment within our 
pixel sampling rather than merely a variation of magnetic field. In regions 
spatially removed from the large flux concentrations of sunspots and plage, 
photospheric magnetic fields are believed to be unresolved kilo--Gauss 
elements, covering a small fraction of the pixel area \citep{lit02, dom03}. In 
such cases the magnetic filling factor, $f$, is much less than unity and 
pixels will contain a contribution from both the ambient high--$\beta$ plasma 
and low--$\beta$ flux concentrations. In the high--$\beta$ material, 
magnetoacoustic slow modes have negligible density variations and are 
unobservable to our broad UV passbands, while the compressible fast modes will 
propagate between $c_S$ and $\sqrt{c_S^2 + v_A^2}$ ($\approx c_S$ since $\beta 
\gg 1$). In the flux concentrations, slow--magnetoacoustic waves will 
propagate at $c_S$ (since $\beta < 1$) while fast modes propagate between 
$v_A$ and $\sqrt{c_S^2 + v_A^2}$. If these flux concentrations satisfy certain 
requirements they may be thought of as ``thin flux tubes'' \citep{rob85}, in 
which slow modes propagate at the tube speed. Values of $0.2-0.5$ are 
typically used for plasma--$\beta$ in the modeling of flux tubes 
\citep[\eg,][]{has03, has05}, yielding tube speeds of $\sim$$0.9 c_S$ and 
fast--mode speeds between $\sim$$(1.5-2.4) c_S$ and $\sim$$(1.8-2.6) c_S$. In 
this picture, the lowest field strengths arise from small values of $f$ so the 
measured speed is dominated by the ``flux--external'' sound speed, while for 
higher field strengths the filling factor increases and the contribution from 
the ``flux--associated'' speeds becomes more important. 

Eventually, a filling factor of unity is reached and the passbands are formed 
in an homogeneously distributed magnetic environment for greater field 
strengths. The change from inhomogeneous to homogeneous environments ends the 
``flux tube'' supported wave scenario. When $f = 1$, the magnetic field has 
expanded to fill the entire pixel area, creating a ``magnetic canopy'' over 
the previously ``flux--external'' plasma. This marks the transition from mixed 
high-- and low-- $\beta$ to entirely low--$\beta$ environment (\ie, the $\beta 
= 1$ level) and is signaled by the change in trend of phase difference 
gradient at fields of $\vert 100-125 \vert$~G. Although kilo--Gauss magnetic 
elements have been mentioned and field strengths of $\vert 100-125 \vert$~G 
are reported for filling factors of unity, field values are measured at the 
photosphere while $f$ corresponds to the region of the chromosphere where 
waves are detected and field lines have expanded into a greater spatial area.

Inside the subsequent homogeneous low--$\beta$ atmosphere, the fast--mode 
propagates at $v_A > c_S$, tending to $\gg$$c_S$ when $\beta \ll 1$. The 
initial fast--mode speed in the homogeneous situation (\ie, when $\beta = 1$) 
is expected to be at most $\sim$$\sqrt{2} c_S$, yielding a phase difference 
gradient which differs from the field--free case by a factor of $\sim$$0.7$ 
(within the error bars of the ratio between our values recorded at $\vert 
100-125 \vert$~G and $\vert 0-20 \vert$~G). The slow mode is again 
quasi--longitudinal and propagates at $c_S$. We note that the sound speed 
should not change between field strengths since it depends only on temperature 
($\propto$$\sqrt{T}$) and the wave behaviours reported here occur between the 
same two UV passbands, which are formed over the same range of temperatures 
irrespective of magnetic field. The gradual decrease of wave speeds in the 
range $\vert 125-550 \vert$~G is taken as a signature of both modes existing 
within our signal, with the measured value of $M_{\Delta \phi}$ resulting from 
averaging between the two speeds. The increase of phase difference gradient 
with increasing magnetic field towards a similar value as detected over field 
strengths in the small filling factor range of $\vert 0-20 \vert$~G implies 
that the wave speeds are tending toward the sound speed, and slow--mode waves 
become dominant in our stronger field regions of plage and umbrae. An 
implication of slow--mode dominance in regions of stronger field is that slow 
modes propagate higher into the atmosphere than fast modes \citep{ost61} --- 
the ``magnetic canopy'' occurs at lower altitudes when moving to higher field 
strengths, the UV passbands are formed progressively further above the 
$\beta$--unity level, and we detect only the speeds of modes which propagate 
higher into the atmosphere. The disclosure of these waves as slow modes 
provides a direct, and necessary, link to coronal studies \citep{demoo02}, 
where propagating intensity variations at 3-- and 5-- min periods are 
interpreted as slow--magnetoacoustic waves in loops anchored in umbrae and 
plage, respectively. We note a disparity between the abundance of wave 
signatures in the low chromosphere and the sparse detection of slow--mode 
associated phenomena in the transition region \citep{depon04} and corona 
\citep{demoo02}. Some mechanism is clearly required to quench most of these 
lower atmospheric waves with one possibility being slow--mode driven Pedersen 
current dissipation \citep{goo00}, theorised to be highly efficient in 
the chromosphere. The non--uniformity observed in the upper atmosphere may 
highlight other selection effects at work on wave transmission, possibly 
based on $p$--mode interference with complex field orientations. Field 
inclinations affected by $p$--modes may also help explain the temporal 
intermittency of waves by only providing suitable propagation conditions over 
short, repeating timescales.

It should be noted that some penumbral pixels may be misinterpreted as weaker 
field regions from of the LOS nature of \mdi\ and the existence of 
highly--inclined, outer--penumbral magnetic fields, \eg, field strengths of 
$\vert 800-1000 \vert$~G inclined 80\degr\ to the vertical \citep{sol92} will 
be recorded as $\vert 140-175 \vert$~G. However, pixels misclassified as such 
will not greatly modify the phase difference gradients, as the true 
weak--to--moderate field pixels ($\vert 25-550 \vert$~G) dominate the number 
statistics. Instead they contribute to greater error--bar magnitudes at higher 
field strengths, alongside the decrease in pixel numbers. Larger errors 
associated with gradient values above $\vert 550 \vert$~G may arise from 
averaging between the sunspot $\sim$3-- and 5-- min distributions, if 
propagating at different speeds. Decreased gradient values over $\vert 550-950 
\vert$~G from those at fields both above and below complements this 
interpretation, as period PDFs exhibit widening toward and the detection of 
5--min penumbral oscillations over this range. Hence, by causing decreased 
phase difference gradients, penumbral waves appear to either travel faster 
than those in umbrae and plage or experience a reduction in the HOF 
separations, distinctly possible for a penumbral atmosphere stratified along 
field lines at inclinations of $40-70$\degr\ \citep{mat03}.

\section{CONCLUSIONS}
\label{sec:conc}
Chromospheric intensity variations have been studied using both 
time--localized wavelet and Fourier analysis. Investigation of the oscillation 
behaviours has identified three regimes with distinctly differing properties.

The increase of wavelet period with \mdi\ field over $\vert 25-100 \vert$~G 
marks the modification of the magnetoacoustic cutoff period by field 
inclination effects. This arises from the ``magnetic canopy'' being pulled 
progressively closer to the sampled HOFs so that they experience more inclined 
fields. Wavelet durations show a decrease in the 1550~\AA\ data over those in 
the 1600~\AA\ and 1700~\AA\ data, indicating a loss of wave energy. This can 
be explained by waves undergoing mode conversion or reflection at the 
$\beta$--unity surface \citep{bog03}. Fourier analysis indicates that wave 
phase speeds increase with magnetic field between the 1700~\AA\ and 1600~\AA\ 
formation heights. It is proposed here that this is due to an increase in 
magnetic filling factor within the pixel sampling, yielding a greater 
contribution to the observed wave speed from the low--$\beta$ 
``flux--associated'' fast--mode speed over the slower sound speed of the 
surrounding high--$\beta$ plasma.

The similarity of wavelet periods and durations over field strengths in the 
range $\vert 100-450 \vert$~G implies that waves in this regime do not 
experience a drastic change in their environmental topology as evidenced by 
the weaker--field case. The $\beta$--unity surface is believed to lie below 
the formation height of 1700~\AA\ emission, no transitional interface exists 
between the passbands, and wave energy (\ie, wavelet duration) is conserved 
with altitude. In contrast to the weak--field regime wave speeds decrease, 
which may be a signature of a change from dominant fast--mode magnetoacoustic 
waves to slow--mode waves when moving from network into plage.

Wavelet period distributions broaden and display a doubly--distributed nature 
over $\vert 550-950 \vert$~G, marking the sampling of sunspot regions. 
Interpretation of the 5--min component as being of penumbral origin is given 
strong support by the virtual disappearance of these higher--period components 
when sampling field strengths $\ge$$\vert 950 \vert$~G (\ie, umbrae). The 
decrease of phase difference gradient values over the mixed penumbral/umbral 
field strength range ($\vert 550-950 \vert$~G) suggests that either penumbral 
waves travel at greater speeds than those in plage and umbral regions or the 
formation heights of the 1700~\AA\ and 1600~\AA\ passbands are closer in 
altitude (\eg, because of inclined fields) in the penumbra.

In the future, we will be able to better determine the intricate relationship 
between magnetic topology and propagating waves using high--resolution, vector 
magnetic field information. Such data will soon be obtained by the 
full--Stokes spectropolarimeter in the instrument suite of the Solar Optical 
Telescope ({\sc{sot}}) Focal Plane Package ({\sc{fpp}}) on board 
{\sl{Solar--B}}. Coordinated observations between this instrument, \trace, and 
the multi--wavelength tunable filters also contained in the 
{\sc{sot}}/{\sc{fpp}} will yield diffraction--limited, high--cadence imaging 
at multiple heights through the outer Solar atmosphere unparalleled by 
previous observations. The combination of these forms of data with a range of 
atmospheric models taking into account the differences in atmospheric 
stratification for differing magnetic fields will yield a powerful diagnostic 
suite for studies of solar wave phenomena.

\acknowledgments
The authors thank the referee, Scott McIntosh, for comments which improved the 
paper. This work was supported by the UK Particle Physics and Astronomy 
Research Council and a travel grant from ESA/\soho. JMA is funded by a 
National Research Council Research Associateship, while FPK is grateful to AWE 
Aldermaston for the award of a William Penney Fellowship.

\end{document}